\documentclass[12pt]{iopart}

\usepackage{graphicx}
\usepackage{multirow}
\expandafter\let\csname equation*\endcsname\relax
\expandafter\let\csname endequation*\endcsname\relax
\usepackage{amsmath}
\usepackage{dcolumn}
\usepackage{subcaption}

\begin{document}

\title{A machine learning accelerated inverse design of underwater acoustic polyurethane coatings with cylindrical voids}

\author{Hansani Weeratunge$^{1}$, Zakiya Shireen$^{1}$, Sagar Iyer$^{1}$, Richard Sandberg $^{1}$, Saman Halgamuge$^{1}$, Adrian Menzel$^{2}$, Andrew Phillips$^{2}$, Elnaz Hajizadeh$^{1*}$ }

\address{$^1$ Department of Mechanical Engineering, Faculty of Engineering and Information Technology, The University of Melbourne, Parkville,
Australia}
\address{$^2$ Maritime Division, Defence Science and Technology Group, Melbourne, Australia}
\ead{ellie.hajizadeh@unimelb.edu.au}

\begin{abstract}Here, we report the development of a detailed ``Materials Informatics" framework for the design of acoustic coatings for underwater sound attenuation through integrating Machine Learning (ML) and statistical optimization algorithms with a Finite Element Model (FEM). The finite element models were developed to simulate the realistic performance of the acoustic coatings based on polyurethane (PU) elastomers with embedded cylindrical voids. The FEM results revealed that the frequency-dependent viscoelastic behavior of the polyurethane matrix has a significant impact on the magnitude and frequency of the absorption peak associated with the cylinders at low frequencies, which has been commonly ignored in previous studies on similar systems. The data generated from the FEM was used to train a Deep Neural Network (DNN) to accelerate the design process, and subsequently, was integrated with a Genetic Algorithm (GA) to determine the optimal geometric parameters of the cylinders to achieve maximized, broadband, low-frequency waterborne sound attenuation. A significant, broadband, low-frequency attenuation is achieved by optimally configuring the layers of cylindrical voids and using attenuation mechanisms, including Fabry-P$\acute{e}$rot resonance and Bragg scattering of the layers of voids. Integration of the machine learning technique into the optimization algorithm further accelerated the exploration of the high dimensional design space for the targeted performance. The developed DNN exhibited significantly increased speed (by a factor of $4.5 \times 10^3$) in predicting the absorption coefficient compared to the conventional FEM(s). Therefore, the acceleration brought by the materials informatics framework brings a paradigm shift to the design and development of acoustic coatings compared to the conventional trial-and-error practices.
\end{abstract}

\vspace{2pc}

\section{\label{sec:level1}Introduction}
	Acoustic coatings are often employed on maritime platforms to reduce underwater noise transference and absorb external acoustic waves \cite{Jayaku2}. Acoustic vibration or sound is a pressure wave transmitted in the form of energy through air, water, or other media. Therefore, attenuation of sound requires reduction or removal of this transmitted energy generally by converting it into heat. Among a range of attenuating materials, elastomeric polymers have been the material of choice for this purpose due to their light weight, lossy and viscoelastic behavior \cite{Jayku}. Furthermore, polymers are easily processed and show good environmental resistance. 
	
	Conventionally, the most commonly used intrinsic sound attenuation mechanism is the absorption by the elastomeric matrix through conversion of sound energy to heat by molecular chain relaxations of the viscoelastic polymers. However, effective attenuation of sound in the low frequency range has been a challenge due to the corresponding long wavelengths, as it involves impractically thick coating requirements. Over the years, several mechanisms such as scattering by redirection, inhomogeneities, and mode conversion \cite{Jayku}, have been developed to improve sound attenuation at broadband low frequencies \cite{Ting}. 
	In low frequency range, sound attenuation has also been achieved by multi-layered composite materials with different natural resonant frequencies in each layer \cite{SHI2019}, by creating periodically distributed voids in the viscoelastic matrix (Alberich-type coatings) \cite{Ivan2} and by embedding soft and hard inclusions within the polymer matrix \cite{FU2021, Xisen}.
	Further, sound wave manipulation using acoustic metasurfaces in the form of thin and lightweight structures is also a central topic in material physics. However, in our study, the focus has been on thick acoustic tiles instead of thin panels. Interested readers on this topic could access the following review paper and references thereof \cite{Assouar2018}. 
	
	Furthermore, various kinds of embedded inclusions such as disk cavities \cite{Disk, Disk2}, cylindrical voids \cite{FEM2, Ivan1, GAO2020107500}, hard spheres \cite{MENG2012, YU2020}, ellipsoidal \cite{elips}, super ellipsoidal inclusions \cite{Ivan2} and combination of both voids and hard inclusions \cite{Sharma1} have been widely examined in polymer matrix coatings. In the detailed study of sound absorption of a viscoelastic coating with steel cylinders and cylindrical void inclusions in a soft rubber medium, it was found that steel and void cylinders result in high sound absorption \cite{Sharma1}. 
	
	Generally, the design of acoustic materials is achieved through employing a \textit{forward approach}, which involves an iterative evaluation of designed and developed coatings through computer simulations or laboratory experimentation. The iterative forward process of predicting the acoustic performance of these coatings to achieve a desired performance is time consuming and computationally expensive, particularly for a high dimensional design space. Nevertheless, the fundamental problem in material design for a specific performance demands an inverse approach, i.e., finding the corresponding design parameters which delivers a \textit{targeted performance}. In the inverse design approach, one can utilize the forward predictive models to compare the performance of the designed materials with the target performance (ground truth). Therefore, the inverse process can be formulated as an optimization problem with an attempt to minimize the difference between the models developed through the forward approach and the ground truth \cite{Zhong2019}. Traditionally, this gap is minimized using empirical trial-and-error methods in conjunction with a prior domain knowledge. With technological advancements and increased computational power, efficient optimization algorithms and data-driven techniques, such as machine learning (ML) can be employed to automate the learning process while utilizing the physics-based understanding \cite{Bacigalupo2020, Donda_2021, Ahmed2021, sun2021acoustic} through physical models.
	
	In the forward approach, analytical and numerical models have been developed to evaluate the acoustic performance of inclusions in a variety of mediums. However, analytical models are usually limited to simple cases, including assumptions with respect to the viscoelastic nature of the polymer matrix and distribution of the inclusions \cite{FEM2, Leroy}. On the other hand, numerical models can be used as an alternative approach for efficient prediction of the acoustic performance of a general periodically voided coating. The finite element method (FEM) \cite{FEM2, FEM1} and layer multiple scattering theory (LMST) \cite{GA_sph} are identified as the prime methodologies for predicting the acoustic performance of these coatings \cite{MENG2012}.

	To reverse engineer acoustic coatings with voids to achieve desired acoustic performance, metaheuristic optimization techniques such as genetic algorithms (GA) \cite{MENG2012, GA_sph, Chang1, Wang_2021} and evolutionary algorithms (EA) \cite{Romero, ZHAO2018} are widely used. Particular interest has been given to GA as it can be used to find a near-optimal solution efficiently \cite{MENG2012, GA_sph, Chang1}. It is reported that the geometric parameters of locally resonant sound scatterers can be optimized to maximize the low frequency sound absorption by a genetic and general non-linear optimization algorithms \cite{MENG2012}. Genetic algorithms have also been used to optimize the parameters of a material with four layers of spherical local resonators to enhance sound insulation by an impedance imbalance and gap coupling \cite{GA_sph}. The performance and quality of the solution of an optimization problem may depend on the choice of the optimization algorithm. Gothall et al. \cite{4_op} compared the performance of various optimization approaches, including adaptive simplex simulated annealing (ASSA), differential evaluation, neighborhood algorithm, and enhanced continuous tabu search, on minimizing the reflection coefficient of an anechoic coating with two layers of spherical cavities. Among the four optimization algorithms, ASSA performed best in terms of speed and the optimal objective function value. They have further observed that some of the optimal parameters reside in the limits of the acceptable range. Thus, the number of dimensions of the problem can be reduced to enhance the performance of the algorithm.
	
	Optimization algorithms often require an iterative evaluation of the objective function. Therefore, computationally efficient modelling approaches are essential in implementing and speeding up the subsequent optimization algorithms. In recent years, the application of Deep Neural Networks (DNN) has become widespread in various fields due to their ability to capture nonlinear dynamics in complex models. Their potential have also been demonstrated in the inverse design \cite{Shi_2020} as well as in accelerated and accurate prediction of the acoustic performance of materials \cite{CIABURRO2021213, JEON2020, IANNA, CIABURRO2020}. The feasibility of Artificial Neural Networks (ANN) was also investigated to estimate the sound absorption coefficient of a fibrous material with four layers \cite{JEON2020}. Application of an ANN has also been shown to predict the acoustic properties of broom fiber considering material properties, geometrical properties, and frequency as inputs \cite{IANNA}.
	
	In the present work we attempt to address the sound attenuation behavior of the acoustic coatings of polyurethane matrix in composite panels with resonant inclusions. We consider simple cylindrical voids as resonators as they couple strongly with water-borne acoustic waves \cite{Leroy}. It is reported elsewhere that, contrary to other shapes of voids, cylindrical voids resonate at lower frequencies \cite{Ivan3, SHARMA2017}. For elastomeric materials, complex dynamic modulus is a key factor to predict accurate damping and absorption capabilities. For this reason we have accounted for frequency and temperature-dependent complex moduli of polyurethane (PU) elastomers in our model by using Dynamic Mechanical Analysis (DMA) data. 
	
	The outline of the study and forward and inverse components of the ``Acoustic Materials Informatics framework" shown in Figure \ref{fig:1} (a) are as follows:

\begin{itemize}
	\item We solve the forward problem by linking the material and geometrical parameters of the polyurethanes and cylindrical voids to their acoustic behavior through a FEM, where we also account for the temperature and frequency dependencies of the Young's and shear moduli of the matrix polyurethanes. 
	\item The Young's and shear moduli master curves are obtained through employing the time-temperature superposition (TTS) principle on experimental data obtained from DMA measurements. 
	\item Data from the FEM is used to develop a deep neural network that efficiently predicts the absorption coefficient of the composites for different design parameters. 
	
	\item In the inverse targeted design component, the neural network model is used in a genetic algorithm to determine the optimal geometrical parameters for each polyurethane matrix composite to maximize broad-band sound absorption at low frequencies.
\end{itemize}

	The manuscript is organized per following sections. Section 2 describes the models developed to establish the acoustic materials informatics framework. Section 2 includes FEM description (section \ref{sec:level2.1}), inverse design using optimization approach (section \ref{sec:level2.2}), followed by accelerated forward approach using deep neural network (section \ref{sec:level2.3}). In section 3, the findings of work are reported in multiple subsections. Section \ref{sec:level3.1} describes the parametric study of the FEM, which includes the effect of frequency-dependent moduli of the PU matrix (section \ref{sec:level3.1.1}). Extension of this parametric study that explores the effect of voids (section S3.1), effect of polyurethane matrix (section S3.2) and finally effect of steel backing (section S3.3) on sound attenuation is presented in the supplementary information. In section \ref{sec:level3.3}, the performances of optimized acoustic coatings with voids in three different polyurethanes (PU80 (section \ref{sec:level3.3.1}), PU65 (section \ref{sec:level3.3.2}) and PU90 (section \ref{sec:level3.3.3})) are compared and discussed in detail. Finally, section 4 summarizes the conclusions emphasising the potential of machine learning approaches in numerical modelling and how the results from this work can be used in further research.
	
\section{\label{sec:level2}Methods and Models : Acoustic Materials Informatics Framework}	
	In acoustic coatings with embedded voids, peak values of the sound absorption coefficient usually correspond to the resonance frequency of the voids. However, the band-gap of this peak may be narrow as it only occurs around the resonance frequency. Therefore, adding multiple voids of different sizes can broaden the band-gap as each of the voids resonates at their natural frequency and results in multiple absorption peaks \cite{FU2021}.  In this study, four cylindrical voids in two different layers with different radii in each layer with varied spacing have been considered per unit cell to create more possibilities for maximizing the absorption coefficient in a wider range of frequencies. 
	
\subsection{\label{sec:level2.1}Finite Element Model (FEM)}
	A 2-dimensional FEM was developed, using acoustic and solid mechanics modules of the COMSOL Multiphysics Software Package version 5.5 to explore the underwater acoustic behavior of a slab of PU matrix with cylindrical voids. We have incorporated frequency-dependent dynamic moduli of the polymer matrix obtained from a DMA (master curves obtained from DMA data is given in Figure S1 (a) in the supplementary information). This allows development of accurate models, which account for the realistic viscoelastic behavior of the polymer matrix and its potential impact on the overall acoustic performance of the acoustic coating.
	
	We considered the cross-section of two layers of cylindrical voids in a PU matrix attached to a steel backing submerged in water as schematically depicted in Figure \ref{fig:1} (b). Periodic boundary conditions were applied to model an infinite array of units in the $y$ direction. The acoustic plane wave was propagated in the $x$ direction and is perpendicular to the $y$ axis. Furthermore, perfectly matching layers were applied at the end of the water layer and air layer to mimic an open and non-reflecting infinite domain. The transmitted pressure was measured at the steel-air interface (the dashed red line), and the reflection was measured at the water-polymer interface (dashed green line). 
	The polymer domain and steel were modelled as viscoelastic solid and elastic solid materials, respectively. The densities of PU and steel were taken as $1026 $ $kg m^{-3}$ and $7850$ $kg m^{-3}$, respectively at temperature $ T= 15^\circ C$ and pressure $P = 101.325$ kPa. The thickness of the steel backing was considered $30$ mm. For air and water, the inbuilt material properties available in the COMSOL materials library were used. The developed FEM models have been validated against data from literature \cite{Jayaku2, Sharma2}, and it is presented in section 2 in the supplementary information. 
	
\subsection{\label{sec:level2.2}Inverse design using optimization approach}	
\begin{figure}
    \centering
	\includegraphics[scale=0.55]{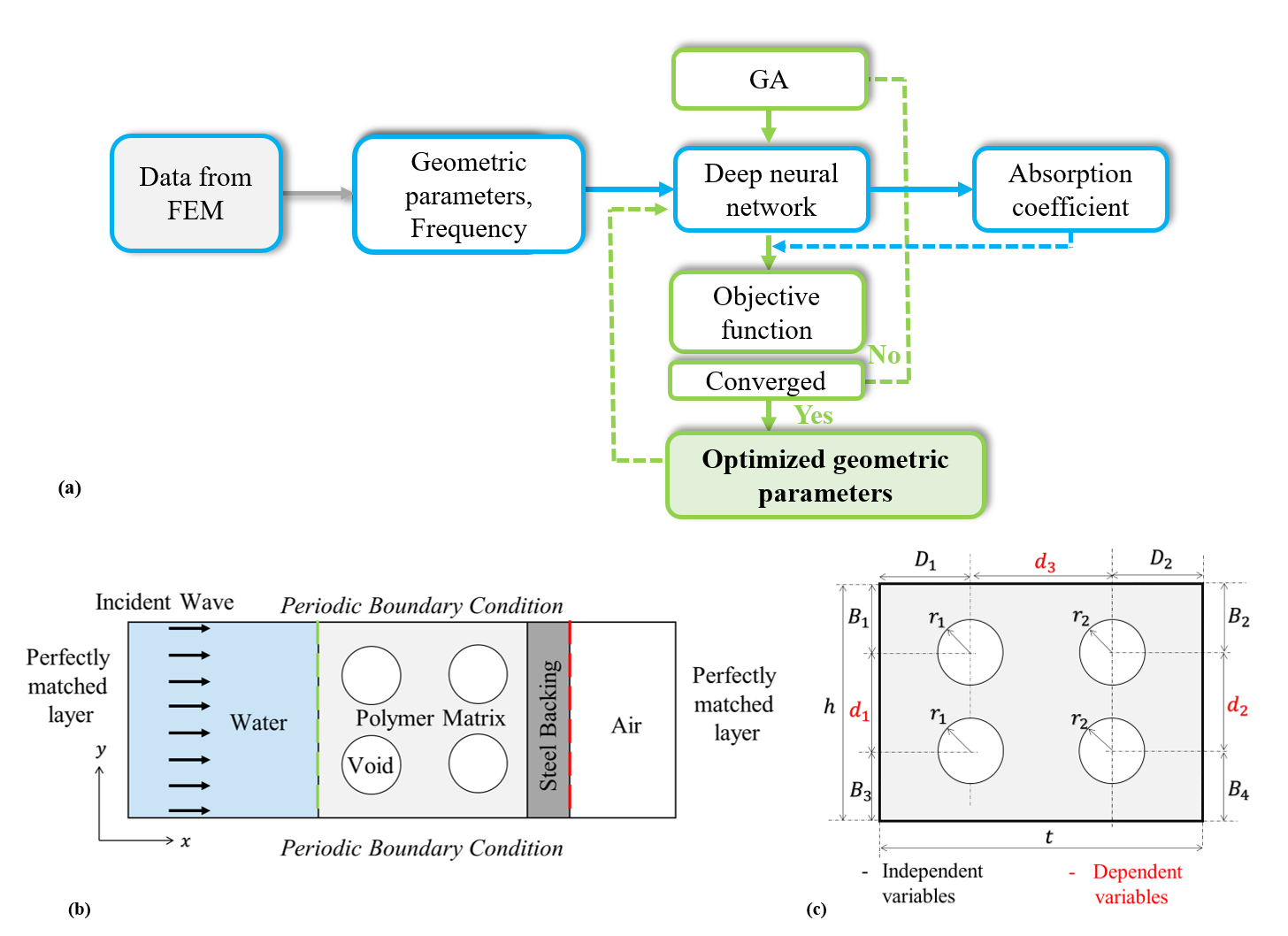}
	\caption{(a) The schematic representation of the proposed materials informatics framework (b) 2D schematic diagram of the unit cell of model polymer matrix composite with four cylindrical voids. (c) Schematic representation of the design space including the geometrical variables for the 2D COMSOL model.}
	\label{fig:1}
\end{figure}

	One of the common and major issues in designing anechoic coatings with embedded voids is the relatively narrow absorption peak produced by the voids \cite{Jayku}. In the present study, we develop an inverse design and optimization framework as shown in Figure \ref{fig:1} (a) to maximize the low-frequency sound attenuation while increasing the bandwidth of the peak. In Figure \ref{fig:1} (b), the schematic of the unit cell of PU matrix composite is shown with four cylindrical voids and is considered for the optimization. An array of two by two voids in the unit cell was considered to account for all the influencing geometrical parameters. This ensures the possibility of exploring a wider design space to find a better solution for a maximized low-frequency broad-band sound attenuation. Furthermore, the design variables of the model are shown in Figure \ref{fig:1} (c). 

	The developed optimization scheme was formulated using the GA as it is an evolutionary-based meta-heuristic search algorithm that mimics the natural selection of a population with the process of adaptation for survival. Genetic algorithm is considered a robust and an efficient approach that can be used to explore complex nonlinear solution spaces \cite{Katoch2021}. Unlike swarm intelligence techniques, GA is less likely to have a premature convergence to a local optimal solution \cite{Katoch2021}. For the optimization algorithm in this study, $10$ independent geometrical parameters of the unit cell (shown in Figure \ref{fig:1} (c)) were considered as the design variables for which objective function is defined as Eq. \ref{eq1},

\begin{equation}
	\begin{aligned}	
	\text{Objective function} = \text{max}({\sum_{i=1}^N}{w_i  a_i - p}),&\\
	{w_i} = \frac{N+1-i}{N}, 
	\label{eq1}
	\end{aligned}
\end{equation}
	
	where $w_i$ and $a_i$ are the weight and absorption coefficient of the $i$th frequency, respectively and $N$ is the number of frequencies. $1000$ frequency points were considered ranging from $10$ Hz-$10$ kHz with equal intervals of $10$ Hz. In the objective function the weighted sum of the absorption coefficient was considered for the whole frequency range but higher weights have been allocated to low frequencies to provide a higher priority to the low frequency range attenuation. Furthermore, a penalty $p$ was included in the objective function to add manufacturing, or practical constraints to avoid infeasible solutions. A constraint was employed to maintain a minimum distance of $1$ cm between the edges of the coating and the voids considering the practical fabrication processes. Larger values of penalty were imposed for solutions that violate the constraints. Additionally, the feasible limits of the design variables were directly incorporated into the GA by setting reasonable lower and upper bounds for the variables presented in Table \ref{table:1}.

\begin{table}[htb]
		\caption{Lower and upper bounds of the design variables}
		\label{table:1}
		\centering
		\begin{tabular}{ |c|c|c|  }
			\hline
			Design & Lower& Upper\\
			Variable &Bound /($mm$) &  Bound/($mm$)\\
			\hline
			$r_1, r_2$ &1  & 15 \\
			$D_1, D_2$ &10  & 80 \\
			$B_1, B_2,B_3,B_4$ &10  & 80 \\
			$h$ &30  & 100 \\
			$t$ &30  & 100 \\	
			\hline
		\end{tabular}
\end{table}

	The transmission coefficient of the acoustic coatings with embedded voids is negligible due to the large impedance mismatch between the steel plate and the air backing (detailed explanation is given in supplementary information S3.3). 
	Therefore, a maximized absorption coefficient is equivalent to a minimized reflection coefficient given their relationship through the following equations 
\begin{equation}
	\begin{aligned}
	A + T + R& = 1, \\
	T & \approx 0, \\
	A + R & =1\\
	\label{eq2}
	\end{aligned}
\end{equation}
	
	where $A$, $T$ and $R$ are absorption, transmission and reflection coefficients, respectively. 
	
Hence, the optimized design of the proposed algorithm will exhibit better anechoic properties with a low reflection coefficient. 

\subsection{\label{sec:level2.3}Accelerated forward approach using Deep Neural Network (DNN)}	
In recent years, the application of artificial neural networks (ANN) is widespread in various fields due to their ability to capture nonlinear dynamics in complex models. Artificial Neural Network is a machine learning technique that mimics the process of the nervous system in the brain. It is used as a basis to develop algorithms with the ability to learn, identify complex patterns, and extract key features from data sets. ANNs are one of the widely used supervised learning models used to develop predictive models in numerous applications.

The most widely used neural network architecture, also known as multi-layer perceptron, consists of multiple layers including an input layer, an output layer, and multiple hidden layers with several nodes per layer. A deep neural network is generally defined as an artificial neural network with more than two hidden layers. The neural networks developed in the present study are DNNs by definition as they consist of multiple layers (greater than two). In a feed-forward neural network, information is passed through from the input layer to the output layer. Information can be propagated to the next successive layer through an activation function that is activated only if the signal values exceed an imposed threshold.  Nonlinear functions such as step, sigmoid or logistic functions are examples of widely used activation functions. However, the choice of activation function is a key factor in designing a neural network as it needs to produce the output within the required range. 

DNN can transform a set of input variables into corresponding outputs by approximating the existing nonlinear relationship. The DNN learns this approximation by `training', i.e., the iterative adjustment of its network parameters to capture the implicit mapping of the input and output variables. This is achieved by utilising an optimisation algorithm to minimise the error between the predicted and actual output. A well-structured and well-trained neural network can accurately predict the output corresponding to a set of previously unseen inputs.

To predict the absorption coefficient of the model shown in Figure \ref{fig:1} (b), a DNN with multiple hidden layers is developed (Note, three separate DNNs were developed to predict the absorption coefficient of the three commercial polyurethanes considered in this study). DNN is an alternate tool to accelerate the forward prediction of the acoustic behavior of the coatings with embedded cylindrical voids. 

The proposed DNN is integrated into the optimization algorithm to expedite the exploration of the best set of parameters that yields low frequency broadband sound attenuation. We employed a multi-layer feed forward neural network with multiple hidden layers as shown in Figure \ref{fig:ML} (a) to predict the absorption coefficient of the acoustic coatings with embedded voids. The input layer considers $10$ independent geometrical parameters (shown in Figure \ref{fig:1} (c)) as well as the frequency as inputs. For enhanced performance of the network, the inputs were converted to a common scale by normalizing to a value between $(0, 1)$. The number of layers, nodes and the hyperparameters of the network such as the learning rate and batch size were tuned using learning curves.
Furthermore, sigmoidal activation functions were used in the network's intermediate layers to activate the signals in the nodes due to its enhanced performance compared to other activation functions and it returns values between $(0-1)$ that align with the range of the data in this study. The network is optimized using the Adam optimizer, which is a stochastic gradient descent algorithm based on adaptive estimates of lower-order moments \cite{ADAM}. 

\subsubsection{\label{sec:level3.2}DNN performance for coating with voids in PU80}\ 
\
\\
\begin{figure}
    \centering
	\includegraphics[scale=0.7]{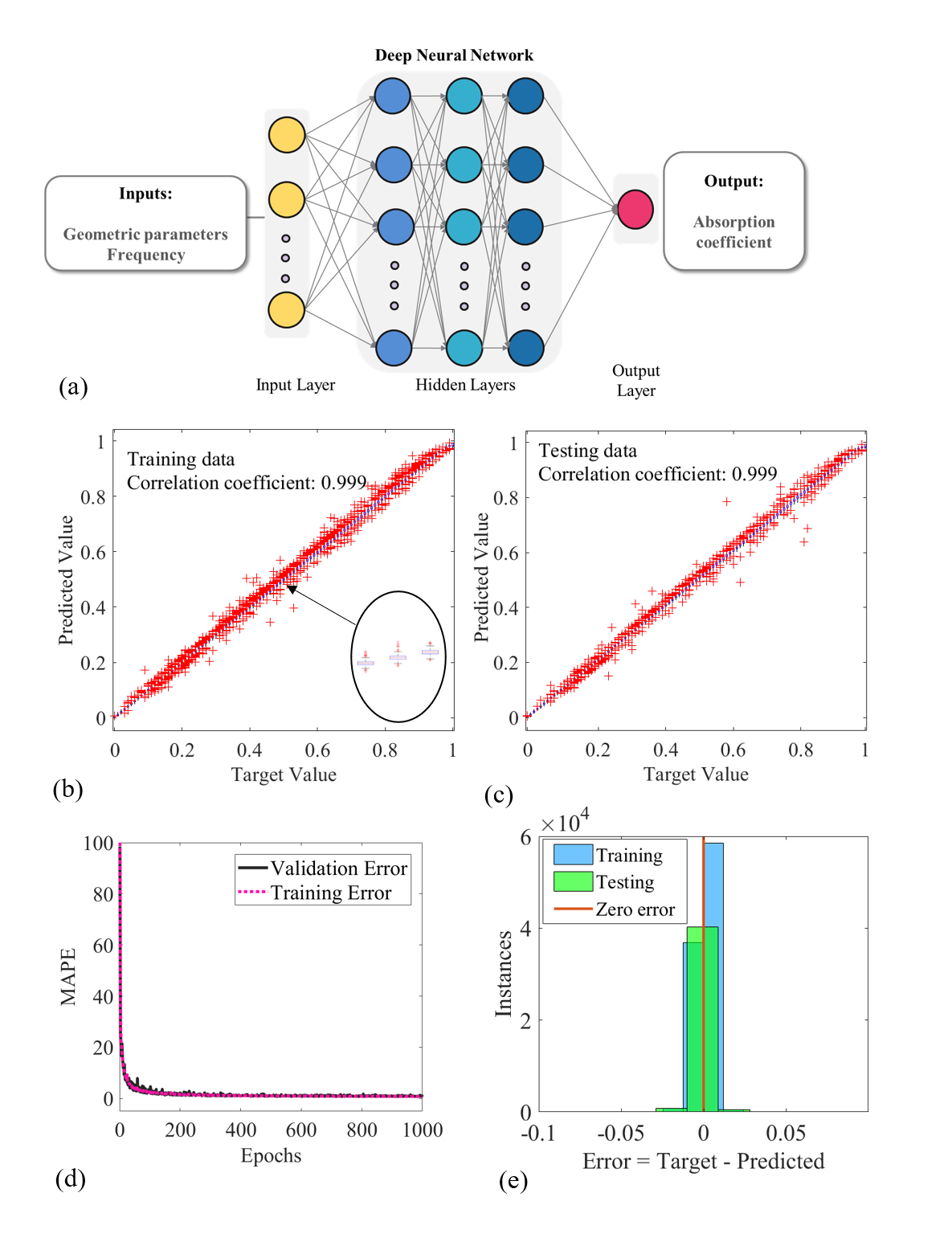}
	\caption{ (a) The general architecture of the DNN developed to predict the absorption coefficient (b) Variation of the predicted (by DNN) and targeted (calculated from the Comsol FEM model) absorption coefficient for the training data (c) Variation of the predicted and targeted absorption coefficient for the testing data. (d) Performance analysis of the neural network during training (e) Error histogram of the neural network. }
	\label{fig:ML}
\end{figure}

	For training the DNN, we collected approximately $150000$ data points from the COMSOL FEM models. This encompasses $400$ sets of combinations of geometric parameters achieved through randomization within the desired range, (presented in Table \ref{table:1}). The simulations were carried out in the low frequency range i.e, $10$ Hz - $10$ kHz. The data were randomly split into two sets. 70\% of the data were randomly selected to train the model and the remaining 30\% of the data are considered for testing. Furthermore, 20\% of the training data were randomly selected for validation. The generality of the network and its performance was evaluated for unseen data. The accuracy of the network was evaluated using the mean absolute percentage error (MAPE) that calculates the deviation between the predicted and actual values. 
	Accurate predictions were achieved with just three hidden layers that contain $200$ nodes (neurons) at each layer. The optimized learning rate, batch size and the number of epochs were 0.0021, 100, and 800, respectively. The convergence of the network during the training process is presented in Figure \ref{fig:ML} (d). Further, the computational time to converge the network from a 2.3 GHz core i7 processor is approximately 820s.
	
	The performance of the trained neural network was investigated by comparing the target value and the predicted value for the absorption coefficient. Figures \ref{fig:ML} (b) and (c) present the distributions of the predictions and the line of equality (identity line) for the training and testing data where the $x$ and $y$ axes are the targets and the predicted values, respectively. The central mark on each blue box represents the median, and the upper and lower edges indicate $25^{th}$ and $75^{th}$ quartiles. The outliers are plotted individually in red `$+$' symbols. The scatter points are closer to the line of equality, which indicates the reliability and accuracy of the prediction. The developed DNN showed a Pearson coefficient (R) of $0.999$ for predicted test data and MAPE of $1.22$, $1.27$, and $1.27$ for the training, validation, and testing data, respectively. Furthermore, as seen in the error histogram in Figure \ref{fig:ML} (e), majority of the predictions are closer to zero error. Therefore, the predicted absorption coefficient from the DNN is in good agreement with the FEM model. Furthermore, the average prediction time for a frequency range from $10$ Hz -$10$ kHz in steps of $20$ Hz for the DNN is approximately $0.04$ seconds, whereas the FEM takes an average of $180$ seconds. This pinpoints the speed increase by a factor of $4.5 \times 10^3$ compared to FEM, and therefore, significantly accelerates the optimization process. 
	
\section{\label{sec:level3}Results and Discussion}
\subsection{\label{sec:level3.1}Parametric study of Finite Element Model}
	We have investigated the impact and physical mechanisms of varying material properties of the polymer matrix as well as geometrical parameters of the cylindrical voids. We also explored the effect of layout of the cylindrical voids on the acoustic performance of the PU80 matrix composite. The detailed discussion is presented in the supplementary information Sections 3.1-3.3.

\subsubsection{\label{sec:level3.1.1}Frequency dependent PU moduli}\ 
\
\\
\begin{figure}
    \centering
	\includegraphics[scale=0.6]{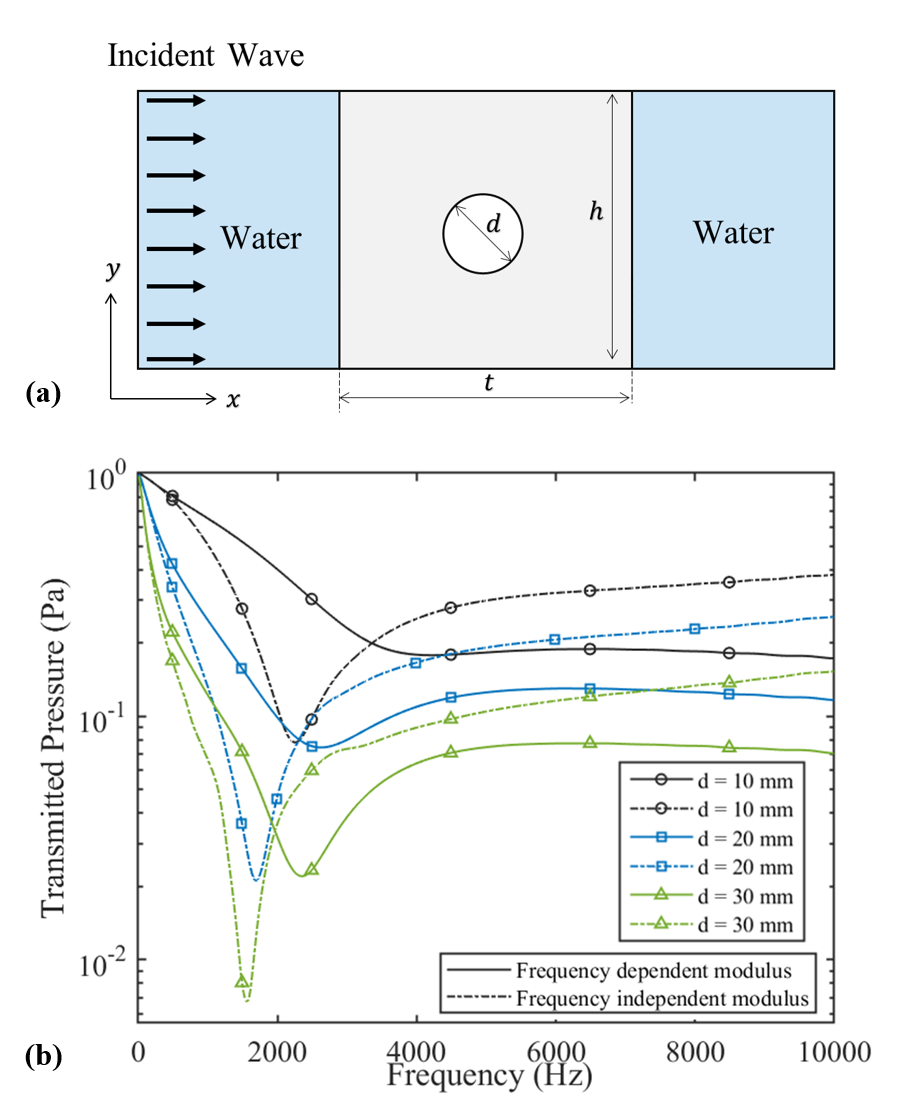}
	\caption{(a) A schematic representation of a model with a single layer of cylindrical voids in PU matrix submerged in water. (b) Transmitted acoustic pressure through PU80 matrix for a range of void diameters. Solid lines represent the frequency-dependent modulus and dashed lines represent the frequency independent modulus.}
	\label{fig:3}
\end{figure}
	In order to consider realistic material properties, frequency-dependent Young's modulus for PU80 (as shown in Figure S1 (a)) was considered. Frequency dependence of sound attenuation of a single layer of voids embedded in the PU matrix (shown in Figure \ref{fig:3} (a)) was evaluated by considering frequency-independent complex Young's modulus at $100 Hz$ ($8.8 + 2.3i$ MPa). The sound attenuation by voids is attributed to their monopole resonance, which results in a reduction in the transmitted acoustic pressure at their resonance frequency \cite{SHARMA2017}. The attenuation is also affected by the void diameter $d$, thus, we varied the void size while keeping the thickness $t = 10$ cm, and the spacing between the voids, i.e., the lattice constant $h = 10$ cm constant, which allowed us to evaluate the effect of the void fraction.
		
	Figure \ref{fig:3} (b) presents the transmitted acoustic pressure of the acoustic coatings for varying diameters, considering frequency-dependent and independent complex Young's moduli. It is observed that there is a significant difference between the monopole resonance frequency, the amplitude, and the bandwidth for the two cases i.e. with (solid lines) and without (dashed lines) frequency-dependent Young's modulus of the PU matrix. The sound attenuation is increased both in amplitude and bandwidth for an increased diameter. This broadband attenuation is attributed to the strong coupling of the voids that corresponds to higher filling fractions. 
	However, the sound attenuation is lower for low diameters due to the reduced coupling between voids. This behavior is observed in both the cases. We have listed the deviations of the monopole resonance for the frequency-dependent and independent modulus in Table \ref{table:2}.

\begin{table}[htb]
\centering
\caption{Monopole resonance for cylinders with different diameters for a polyurethane matrix with and without frequency-dependent Young's modulus and their differences.}
\label{table:2}
\begin{tabular}{ |c|c|c|c|c|  }
\hline
\multicolumn{1}{|c}{}{\multirow{2}{*}{}}&\multicolumn{1}{|c}{Void}{\multirow{2}{*}{}} &\multicolumn{2}{|c|}{Monopole Resonance ($Hz$)} &Percentage\\
\cline{3-4}
$d/t$ &Fraction & Freq. Dependent  & Freq. Independent & Error (\%) \\
 \hline
$0.1$&0.008 &$-$&$ 2250$& $-$ \\
$0.2$&0.031 & $2630$ & $1690$& 35.7\\
$0.3$& 0.070& $2350$ & $1550$& 34.0\\
$0.4$&0.126 & $2290$ & $1610$& 29.7\\
$0.5$&0.196 & $2510$ & $1950$& 22.3\\
$0.6$&0.283 & $3010$ & $2170$& 27.9\\
\hline
\end{tabular}
\end{table}

\subsection{\label{sec:level3.3}Comparison of the performances of different PU matrices}
\subsubsection{\label{sec:level3.3.1}Optimized acoustic coating with voids in PU80}\
\
\\
\begin{figure}
    \centering
	\includegraphics[scale=0.5]{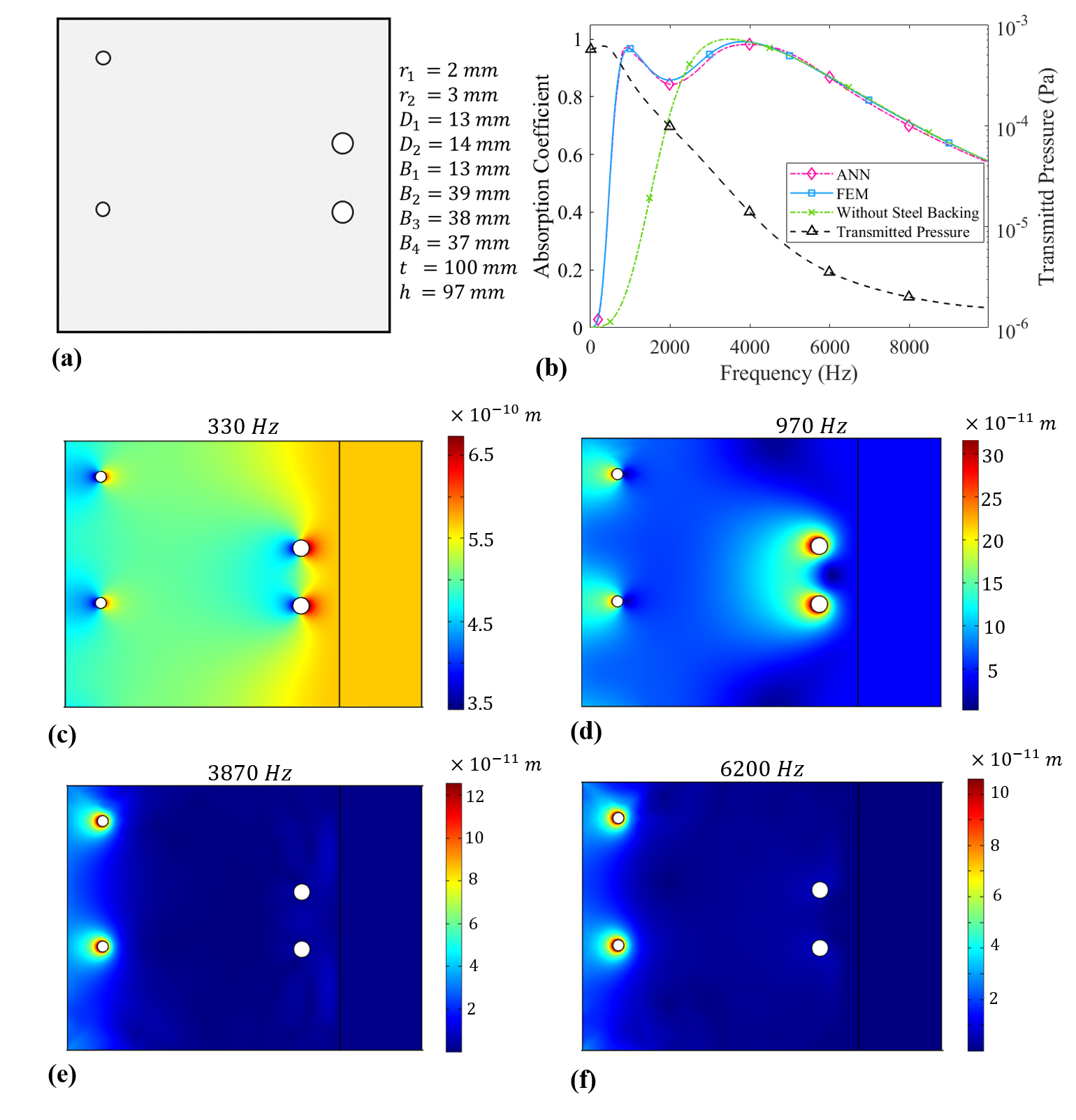}
	\caption{(a) Geometry of the optimized unit cell with PU80 as the matrix material. (b) Broadband sound attenuation of the optimized unit cell of PU80 with embedded voids and steel backing. Deformation maps at (c) $330$ Hz, (d) $970$ Hz, (e) $3870$ Hz, and (f) $6200$ Hz.}
	\label{fig:9}
\end{figure}

	Figure \ref{fig:9} (a) illustrates the geometrical layout of the optimal design that resulted from the developed optimization framework. This design was selected by running the optimization algorithm iteratively and selecting the best option among the solutions as GA does not guarantee the global optimal. The solutions from the iterations converged to similar geometries as shown in the Figure \ref{fig:9} (a) with slight variations in the positioning of the voids. However, the variations in the value of the objective function among these solutions were negligible. Therefore, external factors such as ease of manufacturing and cost can be considered in selecting a suitable solution. Moreover, it was observed during the iterative evaluation that the optimal thickness always resided in the upper bound of the acceptable range, i.e., $100$ mm. This is because having the largest allowed thickness enables attenuation of acoustic waves with a wider range of wavelengths. Thus, the number of dimensions of the problem can be reduced by fixing the values of these parameters. This enhances the efficiency and the speed of the algorithm as the complexity of the problem can be reduced.
	
	Figure \ref{fig:9} (b) shows the absorption coefficient and the transmitted pressure of the optimized acoustic coating with embedded voids in PU80. Absorption coefficient of the acoustic coating with and without the steel backing is further compared. It is observed that the acoustic coating attached to the steel backing exhibits better broad-band absorption characteristics in the low-frequency range. It consists of two peaks at $970$ Hz and $3830$ Hz with absorption coefficients of $0.96$ and $0.99$, respectively. The first peak emerges as a result of the presence of the steel backing at $970$ Hz and is attributed to the scattering of the acoustic wave reflected from the steel backing by the layers of the voids. Also, the highest transmission occurs at $330$ Hz, and its deflection is presented as deformation map in Figure \ref{fig:9} (c). This high transmission corresponds to the high deflection of the steel plate with a magnitude higher than the deflections at other frequencies. Figure \ref{fig:9} (d), presents the deformation of the second layer of voids that interferes with the reflected wave from the steel backing and displays a radial motion as a result of high sound absorption, this finding is consistent with the previous work \cite{motion}. From Figures \ref{fig:9} (e) and f, the first layer of voids that interferes with the incident wave exhibits a radial motion that too corresponds to high absorption. This is further confirmed by Figure \ref{fig:9} (b) that the behavior of acoustic coating with embedded voids in the high-frequency range, i.e. greater than $400$0 Hz is independent of the backing.
	
\subsubsection{\label{sec:level3.3.2}Optimized acoustic coating with voids in PU65}\
\
\\
\begin{figure}
    \centering
	\includegraphics[scale=0.5]{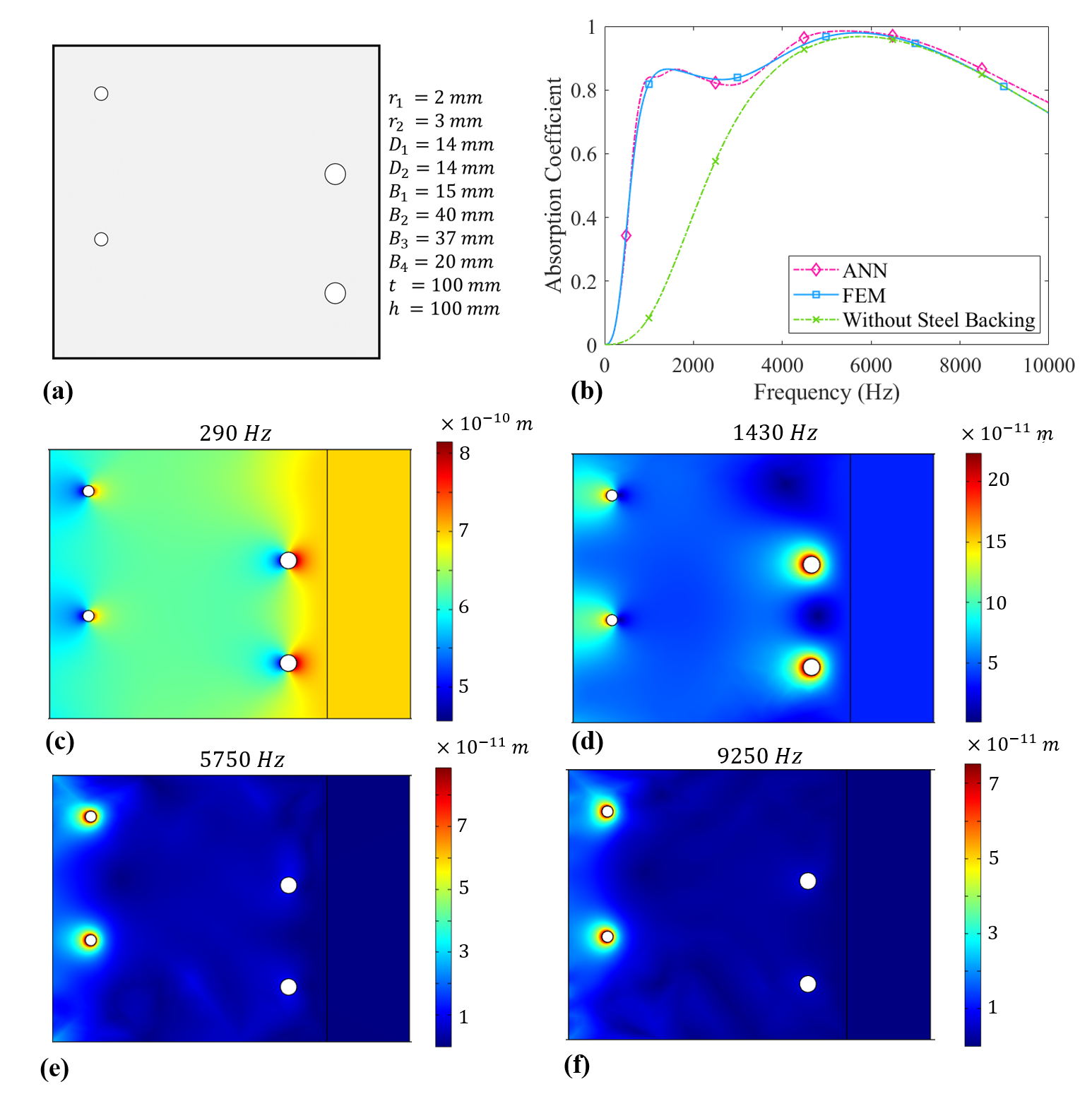}
	\caption{Geometry of the optimized voids in PU65 (a) and absorption coefficient (b). Deformation maps of the optimized voids in PU65 at (c) $290$ Hz, (d) $1430$ Hz, (e) $5750$ Hz, and (f) $9250$ Hz.}
	\label{fig:10}
\end{figure}
	The optimized geometry of the cylindrical voids in PU65 and its performance is shown in Figure \ref{fig:10}, which is rather similar to the PU80 case. As seen in Figure S4 (a) the Young's modulus of PU65 is close to PU80 and from S4b, where both PU65 and PU80 showed low-frequency attenuation with a void of the same diameter. In Figure \ref{fig:10} (b) the absorption coefficient consists of two peaks at $1430$ Hz and $5750$ Hz with coefficients of $0.86$ and $0.98$, respectively. Furthermore, the displacement maps of PU65 at different frequencies are shown in Figure \ref{fig:10} (c, d, e) and (f), where PU65 behavior is observed to be similar to that of the PU80.

\subsubsection{\label{sec:level3.3.3}Optimized acoustic coating with voids in PU90}\
\
\\
\begin{figure}
    \centering
	\includegraphics[scale=0.5]{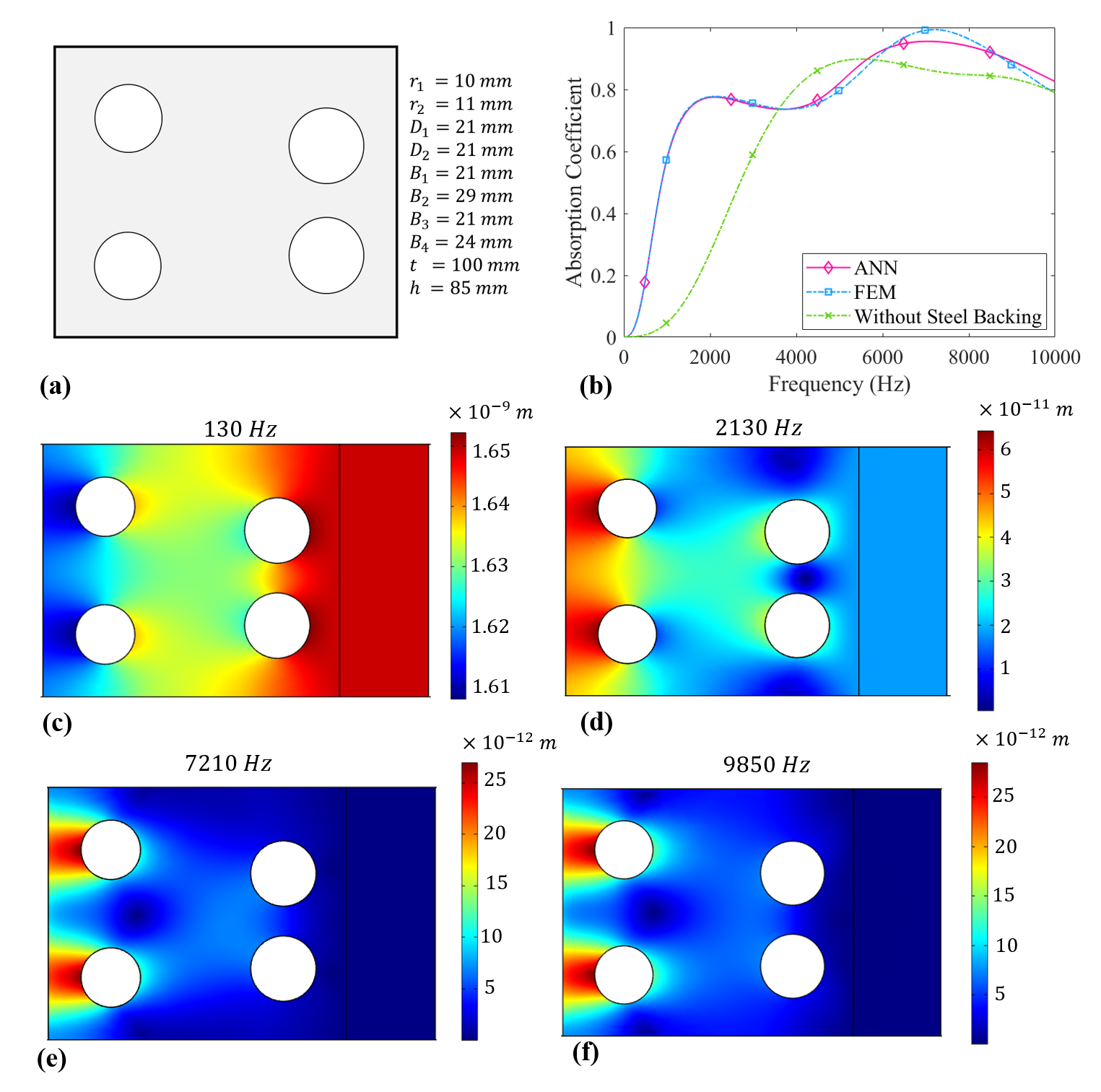}
	\caption{Geometry of the optimized voids in PU90 (a) and absorption coefficient (b). Deformation maps of the optimized voids in PU90 at (c) $130$ Hz, (d) $2130$ Hz, (e) $7210$ Hz, and (f) $9850$ Hz.}
	\label{fig:11}
\end{figure}	
	Figure \ref{fig:11} presents the optimized geometry of the voids in PU90 and its acoustic characteristics. Contrary to PU65 and PU80, the optimal parameters resulted in two layers of voids with larger diameters. PU96 has a higher Young's modulus compared to PU65 and PU80. As shown in section S3.2 an increase in the modulus results in an increase in the speed of sound and therefore, shifts the absorption peaks to higher frequencies. However, as explained in section S3.3, voids with increased diameters can be used to shift the peak to low frequencies to achieve low-frequency attenuation.
	
	The absorption coefficient of the acoustic coating based on PU90 consists of two peaks at $2130$ Hz and $7210$ Hz with coefficients of $0.78$ and $0.99$, respectively. Compared to PU65 and PU80, this corresponds to lower low-frequency sound attenuation. Furthermore, the displacement maps of the unit cell are presented in Figure \ref{fig:11} (c, d, e) and (f) at different frequencies. The highest deflection of the steel plate occurs at $130$ Hz as shown in Figure \ref{fig:11} (c) and this frequency corresponds to the highest transmission. The first peak at $213$0 Hz corresponds to the Fabry-P$\acute{e}$rot resonance, and the regions between the voids of minimum displacement i.e., between the two voids in the right layer correspond(s) to the destructive interference of the sound waves reflected by the steel backing and the voids. At high frequencies, the minimum displacement regions can be seen between the two voids in the left layer that occur mainly due to the interference of the incident wave and sound waves scattered by the four voids. Further, unlike in PU65 and PU80, the effect of the steel backing can be seen throughout the frequency range as shown in Figure \ref{fig:11}. 
		
\section{\label{sec:level4}Conclusion}	
	We developed a materials informatics framework by integrating machine learning and GA optimization algorithms with FEM physical models to develop acoustic coatings comprising cylindrical voids in polyurethane matrix with the aim to achieve maximized, broadband, low-frequency waterborne sound attenuation. A FEM-based physical model was used to simulate the realistic acoustic characteristics of the acoustic coatings with embedded voids through accounting for the temperature and frequency dependent Young's modulus of the three commercial polyurethanes. The data generated from the FEM was then used to train a deep neural network to accelerate forward predictions i.e., the pace of exploring the vast design space. In the inverse design, this neural network was incorporated into GA to determine the optimal geometric parameters with maximum broad-band sound attenuation at low frequencies. The outcomes of this work suggest that the acoustic materials informatics framework, i.e, coupled machine learning and optimization algorithms can be utilized to enable an accelerated inverse design of viscoelastic materials with targeted acoustic performance.
		
	Based on the findings of this study, we made the following conclusions:	
\begin{itemize}		
	\item Acoustic characteristics of coatings with embedded voids depends on various factors, including intrinsic material properties of the polymer matrix (specifically frequency-dependent viscoelastic modulus) as well as size and parameters associated with the geometrical layout of the resonant cylindrical voids. Furthermore, it was observed that temperature and frequency-dependency of the moduli of the polymer matrix needs to be considered to obtain accurate and realistic results for the practical application purposes.
	
	\item A significant, broadband, low-frequency attenuation was achieved by optimally configuring the layers of cylindrical voids and exploiting attenuation mechanisms, including Fabry-P$\acute{e}$rot resonance and Bragg scattering of the layers of voids. 	
		
	\item Optimization algorithms can be utilized for efficient exploration of the design space, which enables accelerated development of acoustic coatings with targeted performance and provides a tool to overcome conventional time and cost-intensive ad hoc trial-and-error forward design approaches.
			
	\item Integration of machine learning techniques into optimization algorithms further accelerates the exploration of the high dimensional design space. The developed DNN exhibited significantly increased speed (by a factor of $4.5 \times 10^3$) in predicting the absorption coefficient compared to the FEM.
		
\end{itemize}

	The significant role of frequency-dependent viscoelastic moduli of the polyurethane matrix on the acoustic performance of these coatings underpins the critical need for establishment of the linkages between polyurethane chemistry (molecular scale) and bulk morphology (mesoscopic scale) to that of the macroscopic viscoelastic behavior \cite{pop}. These could be achieved through integrating the molecular dynamic simulations \cite{MD1, MD2, MD3, MD4} and self-consistent field theory \cite{Fred}  calculations into the continuum models in a unified multiscale materials informatics framework. In addition, to assess the performance of coatings under the operational conditions, the effect of hydrostatic pressure (i.e., operation depth) is being studied and will be the subject of future paper.

\section*{Acknowledgements}
This research is supported by the Faculty of Engineering and Information Technology of the University of Melbourne under the Inter Divisional Strategic Allocation grant number 503195 and partially by the Commonwealth of Australia as represented by the Defence Science and Technology Group of the Department of Defence. 

\section*{Data availability statement}
All data that support the findings of this study are included within the article (and any supplementary files).

\section*{Conflict of interest}
The authors declare that they have no known competing financial interests or personal relationships that could have appeared to influence the work reported in this paper.

\section*{References}	
\bibliographystyle{iopart-num}
\bibliography{references}

\providecommand{\newblock}{}
\begin{thebibliography}{10}
\expandafter\ifx\csname url\endcsname\relax
  \def\url#1{{\tt #1}}\fi
\expandafter\ifx\csname urlprefix\endcsname\relax\def\urlprefix{URL }\fi
\providecommand{\eprint}[2][]{\url{#2}}

\bibitem{Jayaku2}
Jayakumari V~G, Shamsudeen R~K, Ramesh R and Mukundan T 2011 {\em J. Acoust.
  Soc. Am.\/} {\bf 130} 724--730

\bibitem{Jayku}
Jayakumari V~G, Shamsudeen R~K, Rajeswari R and Mukundan T 2019 {\em J. Appl.
  Polym. Sci.\/} {\bf 136} 47165

\bibitem{Ting}
Liu J, Guo H and Wang T 2020 {\em Crystals\/} {\bf 10} ISSN 2073-4352

\bibitem{SHI2019}
Shi K, Jin G, Liu R, Ye T and Xue Y 2019 {\em Results Phys.\/} {\bf 12}
  132--142 ISSN 2211-3797

\bibitem{Ivan2}
Ivansson S~M 2008 {\em J. Acoust. Soc. Am.\/} {\bf 124} 1974--1984

\bibitem{FU2021}
Fu Y, Kabir I~I, Yeoh G~H and Peng Z 2021 {\em Polym. Test\/} {\bf 96} 107115
  ISSN 0142-9418

\bibitem{Xisen}
Wen J, Zhao H, Lv L, Yuan B, Wang G and Wen X 2011 {\em The Journal of the
  Acoustical Society of America\/} {\bf 130} 1201--1208 (\textit{Preprint}
  \eprint{https://doi.org/10.1121/1.3621074})
  \urlprefix\url{https://doi.org/10.1121/1.3621074}

\bibitem{Assouar2018}
Assouar B, Liang B, Wu Y, Li Y, Cheng J~C and Jing Y 2018 {\em Nature Reviews
  Materials\/} {\bf 3} 460--472 ISSN 2058-8437

\bibitem{Disk}
Calvo D~C, Thangawng A~L, Layman C~N, Casalini R and Othman S~F 2015 {\em J.
  Acoust. Soc. Am.\/} {\bf 138} 2537--2547

\bibitem{Disk2}
Sharma G~S, Skvortsov A, MacGillivray I and Kessissoglou N 2020 {\em Applied
  Physics Letters\/} {\bf 116} 041602

\bibitem{FEM2}
Cai C, Hung K and Khan M 2006 {\em J. Sound Vib.\/} {\bf 291} 656--680 ISSN
  0022-460X

\bibitem{Ivan1}
Ivansson S~M 2006 {\em J. Acoust. Soc. Am.\/} {\bf 119} 3558--3567

\bibitem{GAO2020107500}
Gao N and Lu K 2020 {\em Applied Acoustics\/} {\bf 169} 107500 ISSN 0003-682X
  \urlprefix\url{https://www.sciencedirect.com/science/article/pii/S0003682X20306046}

\bibitem{MENG2012}
Meng H, Wen J, Zhao H and Wen X 2012 {\em J. Sound Vib.\/} {\bf 331} 4406--4416
  ISSN 0022-460X

\bibitem{YU2020}
Yu T, Jiang F, Wang J, Wang Z, Chang Y and Guo C 2020 {\em Compos. Struct.\/}
  {\bf 248} 112566 ISSN 0263-8223

\bibitem{elips}
Haberman M~R, Berthelot Y~H and Cherkaoui M 2005 {\em J. Acoust. Soc. Am.\/}
  {\bf 118} 2984--2992

\bibitem{Sharma1}
Sharma G~S, Skvortsov A, MacGillivray I and Kessissoglou N 2019 {\em Appl.
  Acoust.\/} {\bf 143} 200--210

\bibitem{Zhong2019}
Zhong J, Zhao H, Yang H, Wang Y, Yin J and Wen J 2019 {\em Scientific
  Reports\/} {\bf 9} 1181 ISSN 2045-2322

\bibitem{Bacigalupo2020}
Bacigalupo A, Gnecco G, Lepidi M and Gambarotta L 2020 {\em J. Optim. Theory
  Appl.\/} {\bf 187} 630--653 ISSN 1573-2878

\bibitem{Donda_2021}
Donda K, Zhu Y, Merkel A, Fan S~W, Cao L, Wan S and Assouar B 2021 {\em Smart
  Mater. Struct.\/} {\bf 30} 085003

\bibitem{Ahmed2021}
Ahmed W~W, Farhat M, Zhang X and Wu Y 2021 {\em Phys. Rev. Research\/} {\bf
  3}(1) 013142

\bibitem{sun2021acoustic}
Sun X, Jia H, Yang Y, Zhao H, Bi Y, Sun Z and Yang J 2021 Acoustic structure
  inverse design and optimization using deep learning (\textit{Preprint}
  \eprint{2102.02063})

\bibitem{Leroy}
Leroy V, Strybulevych A, Lanoy M, Lemoult F, Tourin A and Page J~H 2015 {\em
  Phys. Rev. B\/} {\bf 91}(2) 020301

\bibitem{FEM1}
Panigrahi S, Jog C and Munjal M 2008 {\em Appl. Acoust.\/} {\bf 69} 1141--1153
  ISSN 0003-682X

\bibitem{GA_sph}
Yuan B, Chen Y, Tan B and Li B 2019 {\em Arch. Acoust.\/} {\bf 44} 365--374

\bibitem{Chang1}
Chang Y~C, Yeh L~J and Chiu M~C 2005 {\em Int. J. Numer. Methods Eng.\/} {\bf
  62} 317--333

\bibitem{Wang_2021}
Wang Y, Zhao H, Yang H, Zhong J, Yu D and Wen J 2021 {\em J. Phys. D: Appl.
  Phys\/} {\bf 54} 265301
  \urlprefix\url{https://doi.org/10.1088/1361-6463/abf373}

\bibitem{Romero}
Romero-García V, Sánchez-Pérez J, García-Raffi L~M, Herrero J,
  Garcia-Nieto~Rodriguez S and Blasco X 2009 {\em J. Acoust. Soc. Am.\/} {\bf
  125} 3774--83

\bibitem{ZHAO2018}
Zhao D, Zhao H, Yang H and Wen J 2018 {\em Appl. Acoust.\/} {\bf 140} 183--187
  ISSN 0003-682X

\bibitem{4_op}
Gothall H and Westin R 2005 {\em Swed. def. res. agcy., Stockh.\/}

\bibitem{Shi_2020}
Shi X, Qiu T, Wang J, Zhao X and Qu S 2020 {\em Journal of Physics D: Applied
  Physics\/} {\bf 53} 275105
  \urlprefix\url{https://doi.org/10.1088/1361-6463/ab8036}

\bibitem{CIABURRO2021213}
Ciaburro G, Iannace G, Ali M, Alabdulkarem A and Nuhait A 2021 {\em J. King
  Saud Univ. Eng. Sci.\/} {\bf 33} 213--220 ISSN 1018-3639

\bibitem{JEON2020}
Jeon J~H, Yang S~S and Kang Y~J 2020 {\em Appl. Acoustics\/} {\bf 169} 107476
  ISSN 0003-682X

\bibitem{IANNA}
Iannace G, Ciaburro G and Trematerra A 2020 {\em Appl. Acoust.\/} {\bf 163}
  107239 ISSN 0003-682X

\bibitem{CIABURRO2020}
Ciaburro G, Iannace G, Passaro J, Bifulco A, Marano A~D, Guida M, Marulo F and
  Branda F 2020 {\em Appl. Acoust.\/} {\bf 169} 107472 ISSN 0003-682X

\bibitem{Ivan3}
Ivansson S~M 2012 {\em J. Acoust. Soc. Am.\/} {\bf 131} 2622--2637

\bibitem{SHARMA2017}
Sharma G~S, Skvortsov A, MacGillivray I and Kessissoglou N 2017 {\em Wave
  Motion\/} {\bf 70} 101--112 ISSN 0165-2125 recent Advances on Wave Motion in
  Fluids and Solids

\bibitem{Sharma2}
Sharma G~S, Skvortsov A, MacGillivray I and Kessissoglou N 2017 {\em J. Acoust.
  Soc. Am.\/} {\bf 141} 4694--4704

\bibitem{Katoch2021}
Katoch S, Chauhan S~S and Kumar V 2021 {\em Multimed. Tools. Appl.\/} {\bf 80}
  8091--8126 ISSN 1573-7721

\bibitem{ADAM}
Kingma D~P and Ba J 2015 {\em 3rd International Conference on Learning
  Representations, {ICLR} 2015, San Diego, CA, USA, May 7-9, 2015, Conference
  Track Proceedings\/} ed Bengio Y and LeCun Y

\bibitem{motion}
Meng H, Wen J, Zhao H, Lv L and Wen X 2012 {\em J. Acoust. Soc. Am.\/} {\bf
  132} 69--75

\bibitem{pop}
Hajizadeh E, Yu S, Wang S and Larson R~G 2018 {\em J. Rheol.\/} {\bf 62}
  235--247

\bibitem{MD1}
Hajizadeh E, Todd B~D and Daivis P~J 2014 {\em J. Rheol.\/} {\bf 58} 281--305

\bibitem{MD2}
Hajizadeh E, Todd B~D and Daivis P~J 2014 {\em J. Chem. Phys.\/} {\bf 141}
  194905

\bibitem{MD3}
Hajizadeh E, Todd B~D and Daivis P~J 2015 {\em J. Chem. Phys.\/} {\bf 142}
  174911

\bibitem{MD4}
Prathumrat P, Sbarski I, Hajizadeh E and Nikzad M 2021 {\em J. Appl. Phys.\/}
  {\bf 129} 155101

\bibitem{Fred}
Paradiso S~P, Delaney K~T and Fredrickson G~H 2016 {\em ACS Macro Lett.\/} {\bf
  5} 972--976

\end{thebibliography}


\providecommand{\newblock}{}
\begin{thebibliography}{1}
\expandafter\ifx\csname url\endcsname\relax
  \def\url#1{{\tt #1}}\fi
\expandafter\ifx\csname urlprefix\endcsname\relax\def\urlprefix{URL }\fi
\providecommand{\eprint}[2][]{\url{#2}}

\bibitem{erapol}
Erapol polymers last accessed: 2021-11-09
  \urlprefix\url{https://www.erapol.com.au/}

\bibitem{WLF}
Williams M~L, Landel R~F and Ferry J~D 1955 {\em J. Am. Chem. Soc.\/} {\bf 77}
  3701--3707

\bibitem{Mott}
Mott P~H, Roland C~M and Corsaro R~D 2002 {\em J. Acoust. Soc. Am.\/} {\bf 111}
  1782--1790

\bibitem{Zakiya21}
Shireen Z, Hajizadeh E, Daivis P and Brandl C 2022 Molecular dynamics study of
  the linear viscoelastic shear and bulk relaxation moduli of
  poly(tetramethylene oxide) (ptmo) (\textit{Preprint} \eprint{2202.08993})
  \urlprefix\url{https://arxiv.org/pdf/2202.08993.pdf}

\bibitem{Maiti2016}
Maiti A 2016 {\em Rheol. Acta\/} {\bf 55} 83--90 ISSN 1435-1528

\bibitem{Jayaku2}
Jayakumari V~G, Shamsudeen R~K, Ramesh R and Mukundan T 2011 {\em J. Acoust.
  Soc. Am.\/} {\bf 130} 724--730

\bibitem{Sharma2}
Sharma G~S, Skvortsov A, MacGillivray I and Kessissoglou N 2017 {\em J. Acoust.
  Soc. Am.\/} {\bf 141} 4694--4704

\bibitem{Leroy1}
Leroy V, Bretagne A, Fink M, Willaime H, Tabeling P and Tourin A 2009 {\em
  Appl. Phys. Lett.\/} {\bf 95} 171904

\end{thebibliography}
\end{document}


\title{A machine learning accelerated inverse design of underwater acoustic polyurethane coatings with cylindrical voids}

\author{Hansani Weeratunge$^{1}$, Zakiya Shireen$^{1}$, Sagar Iyer$^{1}$, Richard Sandberg $^{1}$, Saman Halgamuge$^{1}$, Adrian Menzel$^{2}$, Andrew Phillips$^{2}$, Elnaz Hajizadeh$^{1*}$ }

\address{$^1$ Department of Mechanical Engineering, Faculty of Engineering and Information Technology, The University of Melbourne, Parkville,
Australia}
\address{$^2$ Maritime Division, Defence Science and Technology Group, Melbourne, Australia}
\ead{ellie.hajizadeh@unimelb.edu.au}

\section{Young's and shear moduli master curves of PU matrix}

	In the finite element model, the polyurethane domain was modelled as a viscoelastic material. To employ the frequency-dependent moduli data, the complex shear and Young's moduli were measured experimentally. We considered three commercially available polyurethanes, namely Erapol CC65D, Erapol CC80A, and Erapol CC90A for Dynamic Mechanical Analysis (DMA). For simplicity, we assign the following identifiers for these samples: PU65, PU80, PU90, respectively. The two digits in their identification correspond to the hardness of the PU cast sample as reported by the resin manufacturer \cite{erapol}. However, the detailed analysis presented in the main text only considers PU80 as the material of interest since it has better acoustic characteristics compared to PU65 and PU90. 
	
\begin{figure}[!ht]
	\centering
		\includegraphics[scale=0.36]{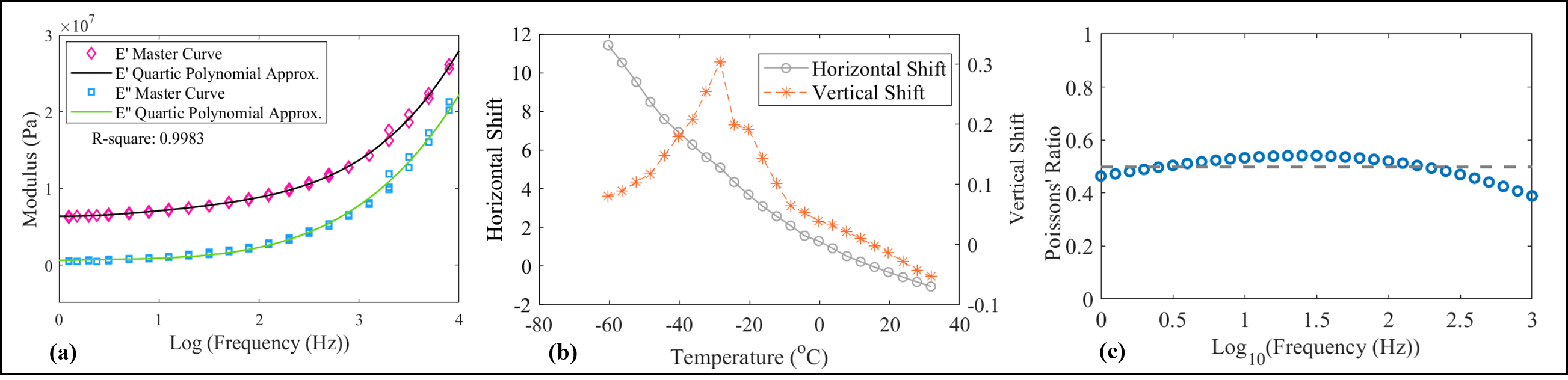}
		\caption{ (a) The master curves for the complex shear and Young's moduli of PU80. (b) Horizontal and vertical shift factors obtained from the optimization scheme. (c) Poisson's ratio plotted for PU80 using Eq. \ref{eq1} utilizing tensile and shear data within the range of frequencies of interest.}
		\label{fig:S1}
\end{figure}	

	The frequency-dependent Young's modulus and shear modulus were measured within the temperature range of $-60^\circ C$ to $ 40^\circ C$. The DMA data was used to generate the master curves using the time-temperature superposition principle. Generally, to construct the master curves horizontal and vertical shift factors are computed using Williams-Landel-Ferry equation \cite{WLF, Mott} using either experimental or computational simulation data \cite{Zakiya21}. However, we determined the shift factors by minimizing the sum of the shortest distances between consecutive curves as optimization techniques are reported to be reasonably reliable compared to subjective methods due to their precision and efficiency \cite{Maiti2016}. The constructed master curves along with the shift factors are presented in Figure \ref{fig:S1} (a, b). The master curves were constructed at $15^\circ C$ for PU80. The Young's modulus was incorporated as a fourth order polynomial function of the frequency in the FEM model.

	Poisson's ratio in the frequency range of interest was obtained from Eq. (\ref{eq1}) using the experimentally measured Young's and shear moduli. As seen in Figure \ref{fig:S1}(c), the Poisson's ratio is almost a constant in the low frequency range, i.e., $1$-$1000$ Hz and, therefore, the average value of $0.499$ has been considered as the input to the FEM model. 

\begin{equation}
\label{eq1}
G = \frac{E}{2(1+v)},
\end{equation}
	where $G$, $E$ and $\nu$ are shear modulus, Young's modulus and the Poisson's ratio, respectively.

\section{Validation of the Finite Element Model}
	The developed FEM model was validated by reproducing the numerical and analytical results obtained from the literature \cite{Jayaku2,Sharma2}. Due to scarcity of literature on acoustic performance of underwater polyurethane coatings with cylindrical voids and frequency-dependent moduli for the matrix, the model was validated using reported studies ref.\cite{Jayaku2, Sharma2}. 

\begin{figure}[!ht]
    \centering
	\includegraphics[scale=0.45]{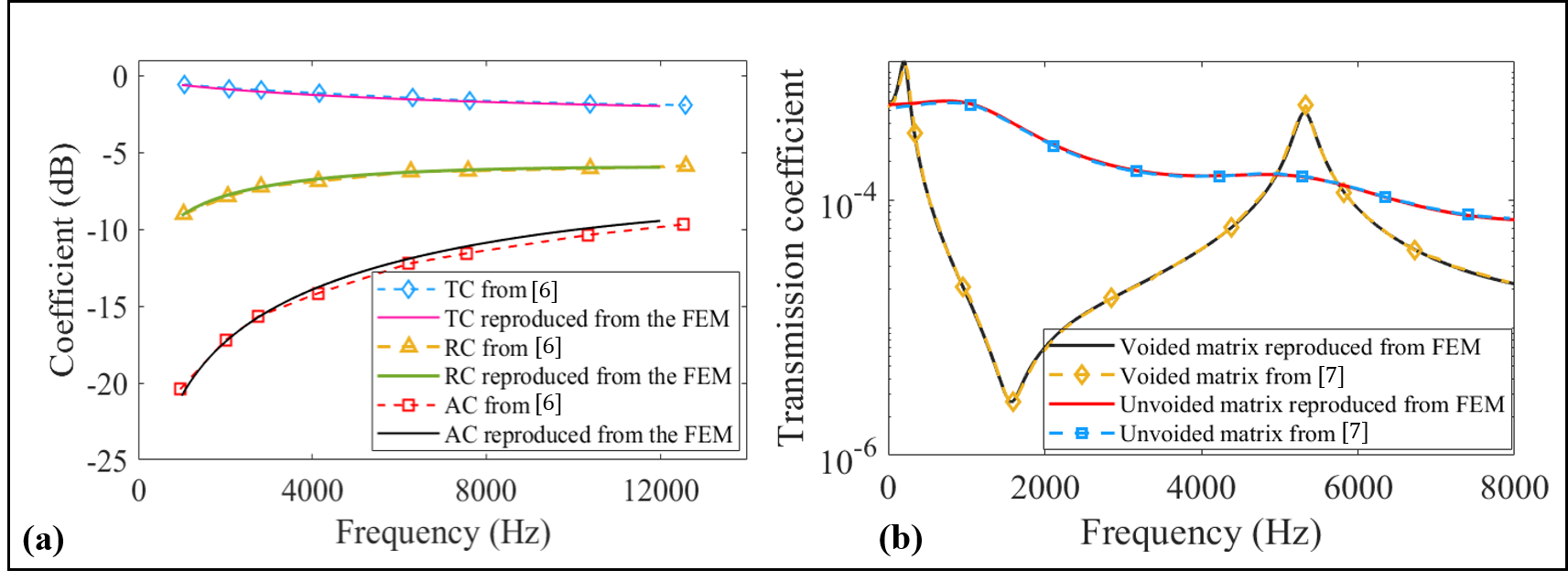}
	\caption{(a) Comparison of transmission (TC), reflection (RC), and absorption (AC) coefficients taken from \cite{Jayaku2} (with permission of the Acoustical Society of America), with acoustic characteristics obtained from the developed FEM, for a panel of pure polyurethane considering frequency dependent Young's modulus.(b) Comparison of Acoustic performance copied from \cite{Sharma2} (with permission of the Acoustical Society of America), and acoustic performance obtained through the developed FEM. Results for the voided medium are compared with the corresponding unvoided medium.}
	\label{fig:S2}
\end{figure}

	To validate the FEM model of a pure panel of PU-based underwater acoustic coating, using the frequency-dependent moduli the data was taken from ref.\cite{Jayaku2}. As seen from Figure \ref{fig:S2} (a), the acoustic behavior of the panel is in good agreement with that obtained from our FEM model, where we have accounted for the frequency-dependent moduli. 

	The developed FEM was used to reproduce the acoustic behavior of a rubber-like medium with cylindrical voids as reported in \cite{Sharma2}. Figure \ref{fig:S2} (b) presents the transmission coefficient of the unvoided and voided matrix cases along with the data from the literature. The acoustic behavior of the coating from our FEM is in good agreement with the literature \cite{Sharma2}. These two comparisons validate the accuracy of the finite element models developed in the present study.  

\section{Parametric study of Finite Element Model}

\subsection{Effect of voids}
\begin{figure}[!ht]
        \centering
		\includegraphics[scale=0.4]{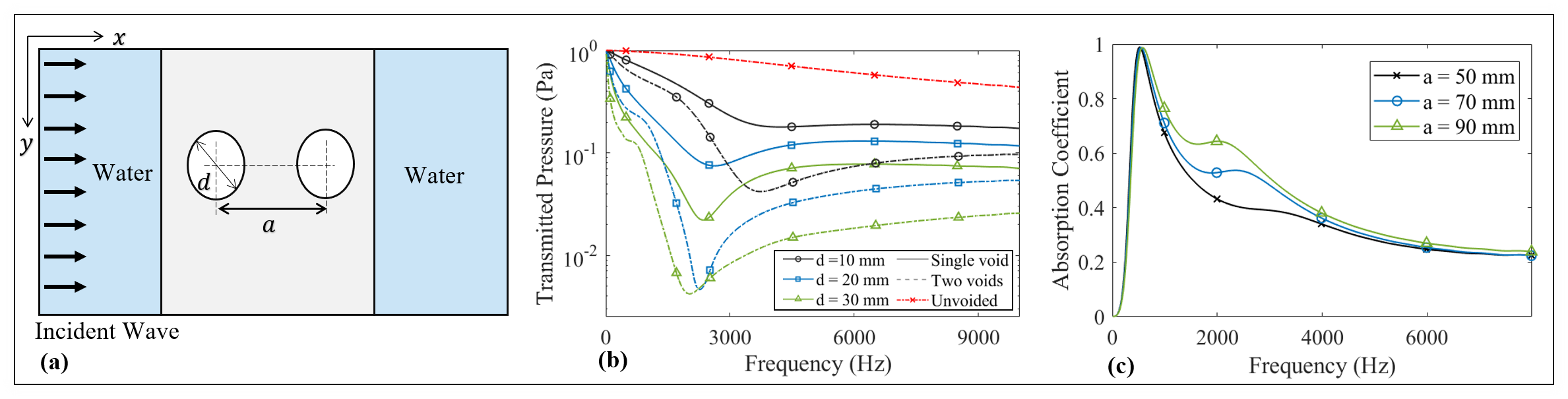}
		\caption{(a) Schematic of the symmetrically distributed voids ($a = 40$ mm). (b) Effect of a single- and double-layer of cylindrical voids with varying diameter on the sound attenuation of the polyurethane matrix. (c) Effect of the distance between the two layers of the voids on the sound attenuation of the PU matrix. }
		\label{fig:S3}
\end{figure}

	Figure \ref{fig:S3}(a) visualizes the schematic of a symmetrically distributed voids in the PU matrix. Figure \ref{fig:S3} (b) represents the comparison of the sound attenuation of the unvoided PU80 matrix, PU matrix with a single layer of voids and two layers of voids with spacing $a$. Figure \ref{fig:S3} (b) also compares the transmitted pressure for voids of varying diameters with the constant ($10$ cm) thickness and the lattice constant of the matrix. The unvoided matrix has a high transmission of the sound wave resulting from its close acoustic impedance with water. However, the presence of voids drastically reduces the transmission around the monopole resonance of the voids and leads to a high sound attenuation over a broader frequency range. Furthermore, it is observed from the Figure \ref{fig:S3} (b) that the two layers of voids have a very high attenuation compared to a single layer of voids. This additional attenuation occurs due to Bragg diffraction and the destructive interference of sound waves \cite{Leroy1}. 

	Figure \ref{fig:S3} (c) presents the absorption coefficient as a function of frequency for different model coating systems, where we vary the distance $a$ between the two layers of voids while keeping the coating thickness constant. We observed that a highly desired secondary low-frequency peak emerges, where an increase in the spacing between the layers results in an increase in the magnitude of this secondary peak and a shift to the lower frequencies. In the proposed optimization framework, this secondary peak is optimized to achieve a broad-band low-frequency behavior for a coating with limited thickness. 

\subsection{Effect of PU matrix}
\begin{figure}[!ht]
	\centering
		\includegraphics[scale=0.4]{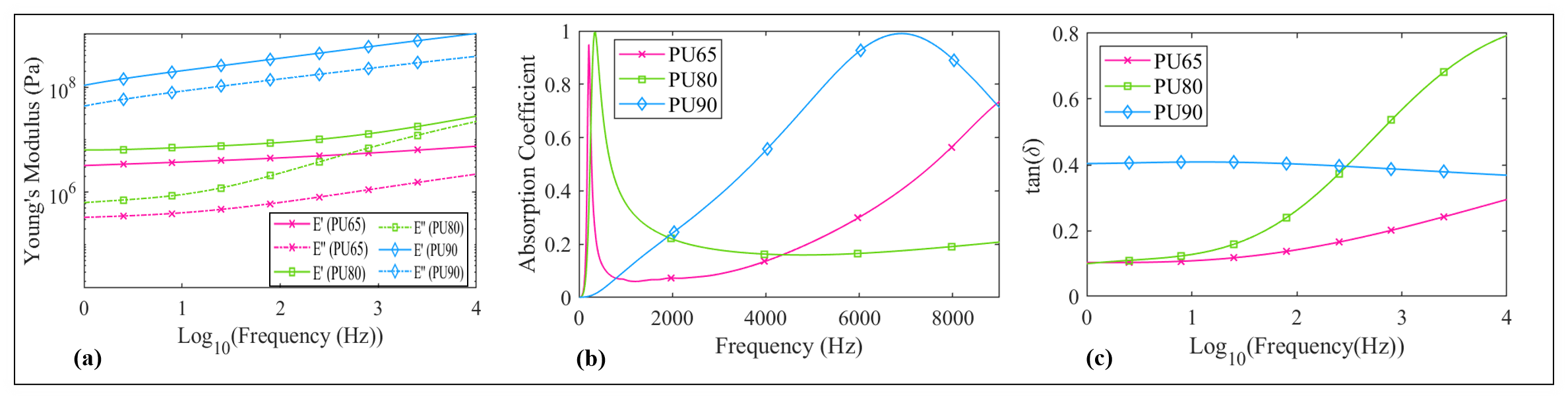}
		\caption{(a) The Young's modulus (b) absorption coefficient (c) and loss tangent tan$\delta$ of model acoustic coatings with three different commercially available polyurethane matrices with a single layer of cylindrical voids.}
		\label{fig:S4}
\end{figure}
	Dynamic Moduli and loss tangent are key factors in determining the acoustic behavior of polyurethanes. We present the sound attenuation of model acoustic coatings with three different commercially available polyurethanes with a single layer of voids (diameter of $20$ mm) attached to the steel plate. The complex Young's modulus and the absorption coefficient of these PUs are presented in Figure \ref{fig:S4}(a) and (b), respectively. The PU65 and PU90 have lowest and highest Young's modulus, respectively. In comparison to the PU80, the hardness is also the lowest in PU65 and highest in PU90. From the Figure \ref{fig:S4} (c), PU80 has overall high loss tangent compared to other two polymers with broader frequency range. Figure \ref{fig:S4} (a) underlines that increasing Young's modulus shifts the first absorption peak to high frequencies and also broadens the bandwidths. Increasing the modulus increases the sound speed in the medium, therefore, shifts the peak to high frequencies \cite{Sharma2}.

    \subsection{Effect of steel backing} 
\begin{figure}[!ht]
	\centering
		\includegraphics[scale=0.45]{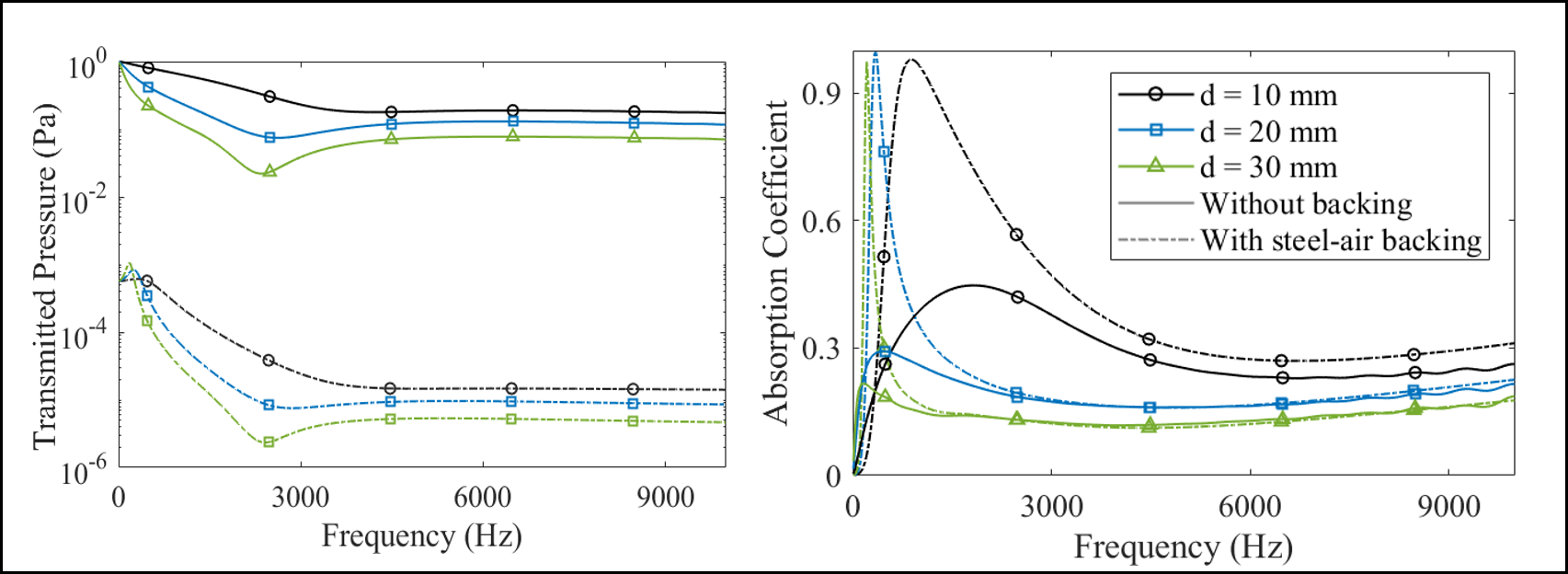}
		\caption{Effect of the steel backing on sound attenuation of a single layer of voids with varying diameters (a) Transmission coefficient and (b) Absorption coefficient. The solid lines represent the coefficients without backing and dashed lines are with steel-air backing. }
		\label{fig:S5}
\end{figure}

    The models developed in this study are based on the assumption that the acoustic coating is generally attached to a structure/vehicle made of steel backed by air (as depicted in Figure 1 (b) of the main text. The presence of this steel-air backing affects the acoustic characteristics of the voids embedded in a viscoelastic polymer matrix. Figure \ref{fig:S5} presents the transmitted pressure and the absorption coefficient of the composite matrix with and without steel-air backing for varying void diameter. 

    As shown in Figure \ref{fig:S5} (a) the presence of the steel-air backing reduces the transmission of sound significantly due to the complete reflection that occurs at the steel-air interface resulting from the high impedance mismatch. However, it is important to note that the resonance frequency of the voids is independent of the type of backing. In addition, a significant increase in sound attenuation through absorption mechanism can also be achieved with the presence of the steel-air backing due to the Fabry-P$\acute{e}$rot resonance mechanism \cite{Sharma2}. Because it results in a constructive interference of the scattered waves from the voids and waves reflected from the steel-air interface, and this further leads to a high absorption arising from the conversion of longitudinal waves to shear waves by the voids.

    As seen in Figure \ref{fig:S5} (b), larger voids with increased diameters shift the first peak of the absorption to lower frequencies, which indicates that the diameter of the void plays a critical role in the acoustic behavior of the composite matrix attached to the steel-air backing. However, the amplitude and the bandwidth are decreased with increasing the diameter due to the high sound reflection from the voids. On the other hand, the low diameters result in higher bandwidth with low amplitudes due to the reduced blocking of reflected waves from the backing. This underlines the significance of selecting an optimal radius to achieve peaks with high amplitudes as well as broader bandwidth at the desired lower frequencies. 
    
\section*{References}
    
\bibliographystyle{iopart-num}
\bibliography{references_SI}